\documentclass[final,3p,times,10pt,nopreprintline]{elsarticle}
\ifdefined\pdfminorversion\pdfminorversion=7\fi

\usepackage[T1]{fontenc}
\usepackage[utf8]{inputenc}
\usepackage{amsmath,amssymb}
\usepackage{graphicx}
\usepackage[table]{xcolor}
\usepackage{booktabs,array,multirow,makecell,tabularx}
\usepackage[ruled,vlined]{algorithm2e}
\usepackage{subcaption}
\usepackage{tikz}
\usetikzlibrary{positioning,fit,arrows.meta}
\usepackage{xurl}
\usepackage[hidelinks]{hyperref}
\hypersetup{
  pdftitle={Rethinking Model Aggregation for Federated Learning with Domain Skew: A Knowledge-Preserving Perspective},
  pdfauthor={Hao-Tian Wu, Xu Ji, Qiheng Li, Li Xiong, Zibin Pan}
}
\definecolor{lightgray}{gray}{0.94}
\newcolumntype{I}{!{\vrule width 1pt}}
\newcommand{\thickhline}{\noalign{\hrule height 1pt}}
\date{}

\begin{document}
\begin{frontmatter}

\title{Rethinking Model Aggregation for Federated Learning with Domain Skew: A Knowledge-Preserving Perspective}
\author[A]{Hao-Tian~Wu}
\author[B]{Xu~Ji}
\author[A]{Qiheng~Li}
\author[A]{Li~Xiong}
\author[A]{Zibin~Pan\corref{cor1}}
\ead{zibinpan@gzhu.edu.cn}
\cortext[cor1]{Corresponding author.}
\address[A]{Cyberspace Institute of Advanced Technology, Guangzhou University, Guangzhou 510006, China}
\address[B]{Guangdong Key Laboratory of Industrial Control System Security, Guangzhou 510555, China}

\begin{abstract}
Domain skew challenges federated learning by requiring a shared model to accommodate distinct visual distributions. Domain-specific experts can capture these distributions separately, creating an opportunity to integrate their representations in a unified predictor. We propose Federated Knowledge-Preserving Learning (FKPL), a two-stage framework that combines domain-oriented expert training with prototype-based classifier calibration. First, a one-shot subspace-based clustering procedure groups clients for intra-cluster training. The resulting expert encoders are concatenated and frozen, preserving each learned feature mapping during global integration. Second, clients interpolate their concatenated features with class-matched domain prototypes and upload the resulting virtual features. The server trains a global linear classifier on these features, concentrating classifier optimization at the server after a single feature upload. FKPL achieves average accuracies of 73.01\%, 67.00\%, and 64.51\% on Office-Caltech-10, PACS, and a four-domain subset of DomainNet, respectively. On DomainNet, its cross-domain standard deviation is 7.50 percentage points, the lowest among the evaluated methods. With the same concatenated extractor, prototype-based calibration improves average accuracy over federated classifier training by 9.99 and 14.94 percentage points on Office-Caltech-10 and PACS. These results identify classifier calibration as an effective way to combine preserved domain representations into a global predictor.
\end{abstract}

\end{frontmatter}

\section{Introduction}
\label{sec1}

Federated learning (FL) enables distributed data owners to train models collaboratively while retaining their raw training samples locally~\cite{DBLP:journals/pami/HuangYSWLDY24}. Its applications include image classification~\cite{conf/eccv/HsuQ020}, medical analysis~\cite{journals/natmi/BerceaWRA22}, and Internet-of-Things systems~\cite{journals/iot/comprs}. A typical classification model combines a feature encoder, which maps an input to a representation vector, with a classifier, which converts that representation into class predictions. How to aggregate these components becomes consequential when clients observe different domains.

\begin{figure}[tbp]
  \centering
  \includegraphics[trim=260 82 300 52,clip,width=0.62\linewidth]{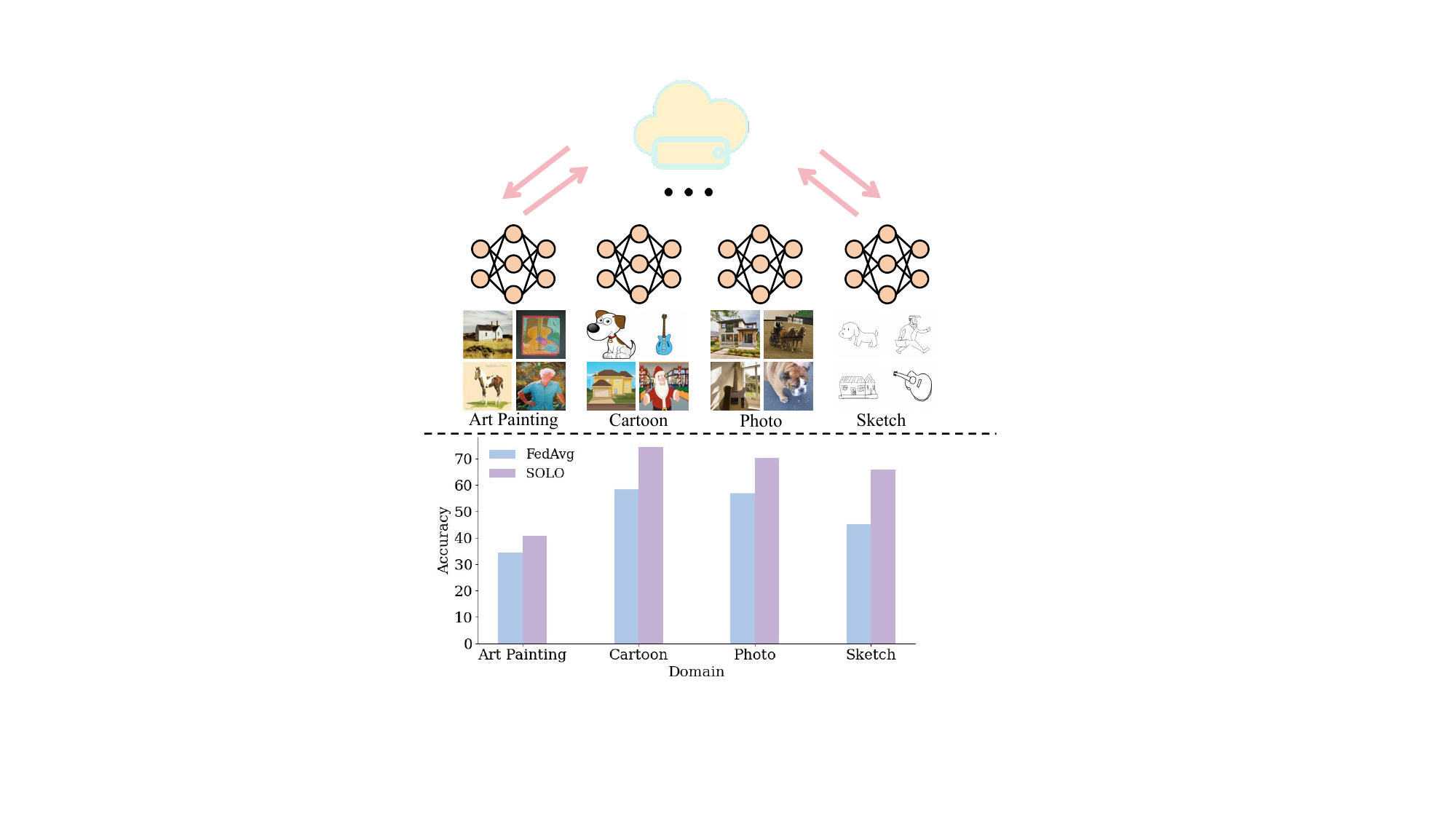}
  \caption{Federated learning across multiple domains. The reported domain accuracies compare FedAvg with domain-specific independent training (SOLO), illustrating the performance gap that motivates preserving specialized representations.}
  \label{fig:problem}
\end{figure}

Data heterogeneity creates differences between local objectives and their optimization trajectories~\cite{journals/ftml/KairouzMABBBBCC21,DBLP:journals/csur/YeFDYT24}. Under domain skew, clients share a classification task while the appearance or acquisition conditions of their inputs vary~\cite{DBLP:conf/cvpr/HuangYSL023,DBLP:conf/cvpr/ChenHY24}. A feature useful in one domain can receive conflicting updates from another domain. Figure~\ref{fig:problem} illustrates the resulting gap between a shared FedAvg model and independently trained domain models. This motivates an aggregation strategy that retains specialized representations while supporting a common prediction interface.

Clustered FL groups clients with similar distributions and trains a model within each group~\cite{journals/tnn/SattlerMS21,DBLP:conf/pmlr/Guo24}. This provides a direct route to domain specialization. Integrating the resulting experts requires an additional mechanism that combines their learned information. Shared-parameter approaches pursue global integration during training through regularization, partial parameter sharing, or modified aggregation~\cite{conf/cvpr/LiHS21,DBLP:conf/cvpr/ChenHY24}. When domain objectives conflict, their shared representations must accommodate competing requirements~\cite{conf/icde/LiDCH22,journals/tkde/ChenDBZJ24}. These observations motivate separating expert training from the construction of the final global classifier.

We propose Federated Knowledge-Preserving Learning (FKPL), following a ``divide-and-integrate'' design. Its first component, \emph{Domain Expert Training and Feature Aggregation} (DETFA), applies one-shot subspace-based clustering~\cite{DBLP:conf/aaai/VahidianMWK0S023}, trains experts within the resulting clusters, and concatenates their feature encoders. Freezing these encoders retains their learned mappings during the integration stage. Every input is then represented by the outputs of all experts, giving a common feature space for global prediction.

The second component, \emph{Prototype-based Classifier Calibration} (PCC), learns how to use this concatenated representation. Clients first compute class prototypes in the frozen global feature space. The server aggregates these prototypes within each cluster and distributes the resulting domain prototypes. Clients then interpolate local features with class-matched prototypes from the represented domains and upload the synthesized virtual features. The server uses this feature collection to train a global linear classifier. This arrangement moves the iterative classifier optimization to the server and makes the classifier's access to cross-domain feature statistics explicit.

Feature concatenation has an established role in FL. FedConcat~\cite{FedConcat} clusters clients using label-distribution information, trains cluster models, concatenates their encoders, and trains a classifier with the encoders frozen. FKPL develops this structural approach for domain-skewed data through two choices: client grouping based on data subspaces and classifier calibration using prototype-guided virtual features. The distinction lies in the grouping information and calibration protocol. Table~\ref{tab:comparison} places these mechanisms alongside representative alternatives.

Experiments on Office-Caltech-10, PACS, and Domain-Net4 examine global prediction, component contributions, representation inversion, and system cost. The strongest component-level gains arise from PCC: with the concatenated extractor fixed, it increases average accuracy by 9.99 percentage points on Office-Caltech-10 and 14.94 points on PACS relative to a federated classifier. FKPL also produces the smallest cross-domain accuracy dispersion on Domain-Net4. These findings support the following contributions:

\begin{itemize}
  \item A domain-oriented integration framework combines subspace-based client grouping, intra-cluster expert training, and frozen encoder concatenation to retain specialized feature mappings in a global predictor.
  \item Prototype-based classifier calibration constructs cross-domain virtual features in a common representation space and trains the global classifier at the server after a single feature-upload phase.
  \item Experiments quantify the gains from classifier calibration, the contributions of expert grouping, and the communication and inference costs of retaining multiple encoders.
\end{itemize}

Section~\ref{br} reviews the relevant FL approaches. Section~\ref{methodology} presents FKPL, Section~\ref{experiments} reports the experiments, and Section~\ref{conclusion} concludes the paper.

\section{Background and Related Work}
\label{br}

\subsection{Federated Learning}
Consider a server and $M$ clients. Client $m$ holds a dataset $\mathcal{D}_m=\{(\boldsymbol{x}_{m,i},y_{m,i})\}_{i=1}^{N_m}$, where $N_m=|\mathcal{D}_m|$. With $N=\sum_{m=1}^{M}N_m$, the standard sample-weighted FL objective is
\begin{equation}
  \min_w \sum_{m=1}^{M}\frac{N_m}{N}\mathcal{L}_m(w),
  \qquad
  \mathcal{L}_m(w)=\frac{1}{N_m}\sum_{i=1}^{N_m}
  \ell\bigl(f(\boldsymbol{x}_{m,i};w),y_{m,i}\bigr),
  \label{eq:objective}
\end{equation}
where $f(\cdot;w)$ is the model and $\ell$ is the per-sample classification loss. FedAvg alternates local optimization with sample-weighted parameter aggregation~\cite{conf/aistats/McMahanMRHA17}.

\subsection{Learning with Heterogeneous Data}
Differences between client distributions can produce local updates that favor different representations~\cite{conf/cvpr/GaoFLC0022}. FedProx constrains local updates using a proximal term~\cite{conf/mlsys/LiSZSTS20}, while MOON regularizes representation learning through a model-contrastive objective~\cite{conf/cvpr/LiHS21}. Such methods modify the optimization of a shared model.

Several approaches explicitly address domain skew. FCCL uses cross-correlation and knowledge distillation to transfer intra-domain and inter-domain information~\cite{journals/pami/HuangYSD24}. FPL uses clustered prototypes to guide representation learning~\cite{DBLP:conf/cvpr/HuangYSL023}. FedHEAL combines partial parameter updates with fairness-aware aggregation to improve performance across domains~\cite{DBLP:conf/cvpr/ChenHY24}. Filter-aware relationships and personalized local structures offer another way to organize heterogeneous representations~\cite{yang2025enhancing}. FKPL retains separate expert encoders and learns a classifier over their combined outputs, making structural preservation part of the aggregation procedure.

\subsection{Clustered Learning and Encoder Concatenation}
Clustered FL organizes clients according to their similarity. Examples use gradient similarity~\cite{journals/tnn/SattlerMS21}, model divergence~\cite{journals/www/LongXSZWJ23}, data-subspace principal angles~\cite{DBLP:conf/aaai/VahidianMWK0S023}, or active-learning indicators~\cite{journals/tnn/HHuang}. PACFL extracts subspace information once before training and uses it to form client groups~\cite{DBLP:conf/aaai/VahidianMWK0S023}. FKPL adopts this type of grouping to train domain-oriented experts.

FedConcat provides a closely related encoder-integration mechanism~\cite{FedConcat}. It groups clients using label-distribution information, trains models within clusters, concatenates the trained encoders, and freezes them during federated classifier training. This establishes encoder concatenation as an existing FL mechanism. FKPL combines subspace-based domain grouping with server-side classifier calibration on virtual features formed from local representations and domain prototypes.

\subsection{Prototype Learning}
A class prototype summarizes feature vectors through a representative centroid~\cite{conf/aaai/TanLLZ00Z22}. FedProto and FedProc use prototypes to guide local training under heterogeneous data~\cite{conf/aaai/TanLLZ00Z22,journals/fgcs/MuSCGFZZ23}, while RUCR uses unified prototypes to address classifier bias under long-tailed distributions~\cite{journals/tifs/HuangLYCD24}. FKPL computes class prototypes separately for each cluster in the same concatenated feature space. These prototypes guide interpolation during classifier calibration, connecting domain-level summaries with sample-level representations.

\subsection{Comparison of Integration Mechanisms}
\label{sec:comparison}
Table~\ref{tab:comparison} compares concrete training and prediction mechanisms. FKPL uses clustering to construct specialized encoders and prototype-guided feature synthesis to fit a shared classifier. Its global prediction evaluates every expert on each input, so the classifier can combine their outputs without selecting a domain-specific model at inference time.

\begin{table}[tbp]
  \centering
  \caption{Representative mechanisms for combining heterogeneous client information. Entries describe the cited methods; prototype use and clustering can also be combined with other FL designs.}
  \label{tab:comparison}
  \small
  \setlength{\tabcolsep}{4pt}
  \renewcommand{\arraystretch}{1.18}
  \begin{tabularx}{\linewidth}{@{}p{0.16\linewidth}*{3}{>{\raggedright\arraybackslash}X}@{}}
    \toprule
    Method & Client or feature organization & Integrated predictor & Classifier training \\
    \midrule
    FedAvg~\cite{conf/aistats/McMahanMRHA17} & Shared parameters & One shared encoder and head & Federated parameter averaging \\
    PACFL~\cite{DBLP:conf/aaai/VahidianMWK0S023} & One-shot subspace-based clusters & Cluster models & Within-cluster training \\
    FPL~\cite{DBLP:conf/cvpr/HuangYSL023} & Prototype-guided representation learning & Shared global model & Federated model optimization \\
    FedConcat~\cite{FedConcat} & Label-distribution-based clusters & Concatenated frozen encoders and one head & Federated training of the global head \\
    FKPL & Subspace-based clusters and class-matched domain prototypes & Concatenated frozen encoders and one head & Server-side training on uploaded virtual features \\
    \bottomrule
  \end{tabularx}
\end{table}

\section{Methodology}
\label{methodology}

FKPL combines domain-specific representation learning with a shared prediction rule. Domain Expert Training and Feature Aggregation (DETFA) learns separate encoders within client clusters and retains their outputs in a concatenated representation. Prototype-based Classifier Calibration (PCC) then trains a global classifier on synthesized features while keeping all encoders fixed.

\subsection{Preliminaries}
\textbf{Problem Statement.} We consider clients that share an input space and a classification task, with data drawn from different domains. Writing the distribution of client $m$ as $P_m(\boldsymbol{x},y)=P_m(\boldsymbol{x}\mid y)P_m(y)$, domain skew refers to differences in class-conditional input distributions. In particular, such differences can arise even when label marginals coincide:
\begin{equation}
\label{eq:domain_skew}
P_i(\boldsymbol{x}\mid y)\neq P_j(\boldsymbol{x}\mid y),
\qquad P_i(y)=P_j(y).
\end{equation}
The participating clients' empirical label proportions can also differ. The learning objective is the global empirical risk in Eq.~\eqref{eq:objective}, and evaluation uses held-out samples from the participating domains. All domains therefore contribute training data to the shared task.

Each client uses a model $f(\boldsymbol{x};w_m)=\mathcal{H}(\mathcal{F}(\boldsymbol{x};\theta_m);\phi_m)$, where $w_m=(\theta_m,\phi_m)$. The encoder $\mathcal{F}(\cdot;\theta_m):\mathcal{X}\rightarrow\mathbb{R}^{d}$ produces a $d$-dimensional representation, and the linear classifier $\mathcal{H}(\cdot;\phi_m):\mathbb{R}^{d}\rightarrow\mathbb{R}^{K}$ produces logits for $K$ shared classes. Clients use the same base architecture and label definitions.

\subsection{Motivation: Preserving Domain-Specific Representations}
\label{sec:motivation}
Parameter sharing and clustered training provide different ways to use heterogeneous data. A shared encoder combines information across clients, while clustered training allows separate encoders to specialize in different distributions. The challenge addressed by FKPL is to combine these specialized representations into one predictor whose parameters can be deployed across the participating domains.

\begin{figure}[tbp]
    \centering
    \includegraphics[trim=155 120 175 60,clip,width=\linewidth]{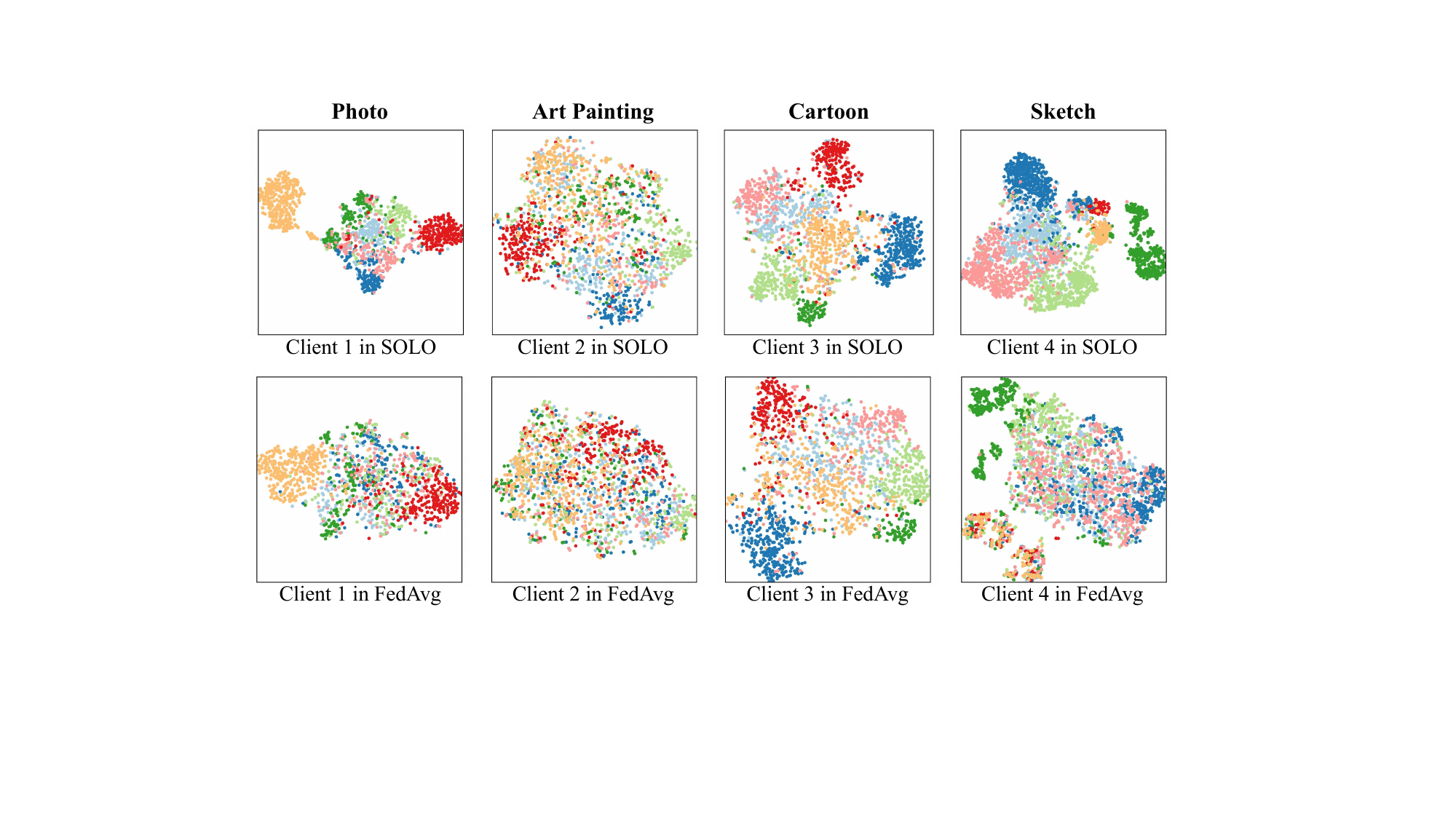}
    \caption{t-SNE visualization of PACS representations. The top row shows independently trained domain models; the bottom row shows the shared FedAvg model. The differences in class separation motivate retaining domain-specific encoders when constructing a unified predictor.}
    \label{fig:motivation}
\end{figure}

An exploratory study on PACS~\cite{DBLP:conf/iccv/LiYSH17} illustrates this motivation. Four clients each hold data from one domain and train a ResNet-10 model~\cite{DBLP:conf/cvpr/HeZRS16}. Figure~\ref{fig:motivation} compares t-SNE embeddings~\cite{van2008visualizing} of penultimate-layer features after 100 epochs of independent training and 100 communication rounds of FedAvg. Independently trained models exhibit clearer class separation in several domains, particularly Sketch. This observation motivates a structural aggregation strategy that retains each trained encoder and learns how to combine its features with those of other domain experts.

\subsection{Proposed Method}
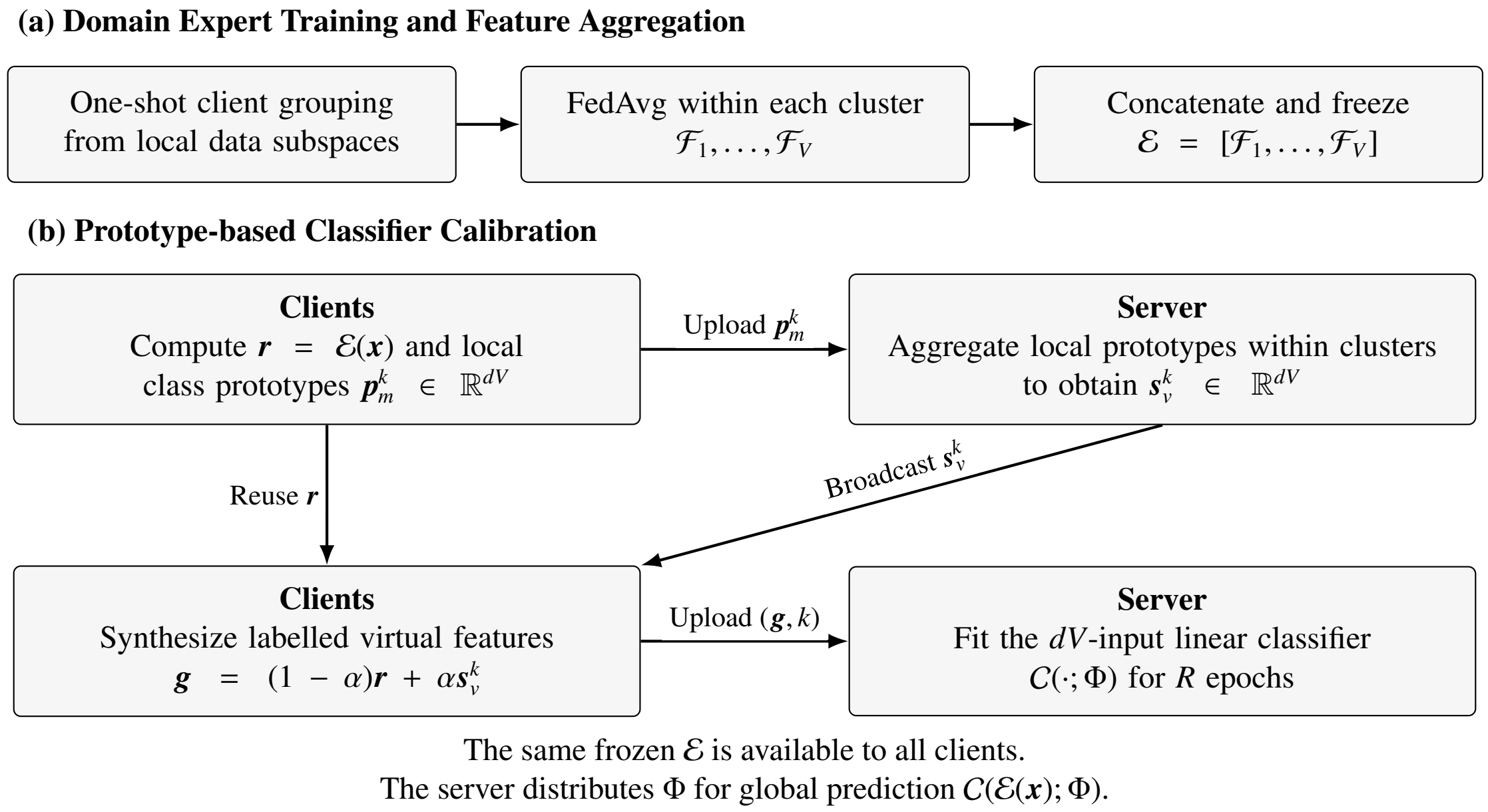
\begin{figure}[tbp]
\centering
\begin{tikzpicture}[
  box/.style={draw,rounded corners=2pt,align=center,font=\small,
              minimum height=1.15cm,text width=4.1cm,inner sep=5pt},
  flow/.style={-{Latex[length=2mm]},thick},
  lab/.style={font=\footnotesize,fill=white,inner sep=2pt}
]
\node[box,fill=gray!7] (group) at (0,0)
  {One-shot client grouping\\from local data subspaces};
\node[box,fill=gray!7] (train) at (5.1,0)
  {FedAvg within each cluster\\$\mathcal F_1,\ldots,\mathcal F_V$};
\node[box,fill=gray!7] (concat) at (10.2,0)
  {Concatenate and freeze\\$\mathcal E=[\mathcal F_1,\ldots,\mathcal F_V]$};
\draw[flow] (group)--(train);
\draw[flow] (train)--(concat);
\node[font=\small\bfseries,anchor=west] at (-2.25,1.0)
  {(a) Domain Expert Training and Feature Aggregation};
\end{tikzpicture}

\medskip
\begin{tikzpicture}[
  box/.style={draw,rounded corners=2pt,align=center,font=\small,
              minimum height=1.3cm,text width=5.8cm,inner sep=6pt},
  flow/.style={-{Latex[length=2mm]},thick},
  lab/.style={font=\footnotesize,fill=white,inner sep=2pt}
]
\node[box,fill=gray!7] (local) at (0,0)
  {\textbf{Clients}\\Compute $\boldsymbol r=\mathcal E(\boldsymbol x)$ and local\\class prototypes $\boldsymbol p_m^k\in\mathbb R^{dV}$};
\node[box,fill=gray!7] (domain) at (8.3,0)
  {\textbf{Server}\\Aggregate local prototypes within clusters\\to obtain $\boldsymbol s_v^k\in\mathbb R^{dV}$};
\node[box,fill=gray!7] (mix) at (0,-2.9)
  {\textbf{Clients}\\Synthesize labelled virtual features\\$\boldsymbol g=(1-\alpha)\boldsymbol r+\alpha\boldsymbol s_v^k$};
\node[box,fill=gray!7] (head) at (8.3,-2.9)
  {\textbf{Server}\\Fit the $dV$-input linear classifier\\$\mathcal C(\cdot;\Phi)$ for $R$ epochs};
\draw[flow] (local.east)--node[lab,above]{Upload $\boldsymbol p_m^k$}(domain.west);
\draw[flow] (local.south)--node[lab,left]{Reuse $\boldsymbol r$}(mix.north);
\draw[flow] (domain.south)--node[lab,sloped,above]{Broadcast $\boldsymbol s_v^k$}(mix.north east);
\draw[flow] (mix.east)--node[lab,above]{Upload $(\boldsymbol g,k)$}(head.west);
\node[font=\small\bfseries,anchor=west] at (-3.1,1.15)
  {(b) Prototype-based Classifier Calibration};
\node[font=\small,align=center] at (4.15,-4.2)
  {The same frozen $\mathcal E$ is available to all clients.\\
   The server distributes $\Phi$ for global prediction $\mathcal C(\mathcal E(\boldsymbol x);\Phi)$.};
\end{tikzpicture}
\caption{The FKPL pipeline. All local prototypes, domain prototypes, and virtual features use the same $dV$-dimensional concatenated representation. Raw images are encoded locally. After virtual-feature collection, classifier optimization is performed at the server with the encoders frozen.}
\label{fig:method}
\end{figure}

Figure~\ref{fig:method} summarizes the two-stage pipeline. DETFA groups clients once, performs federated training within each group, and concatenates the resulting encoders. PCC broadcasts this frozen global extractor, aggregates class prototypes within clusters, and uses same-class prototype interpolation to construct a shared classifier-training set. All prototype and virtual-feature vectors are expressed in the same concatenated feature space.

\subsubsection{Domain Expert Training and Feature Aggregation}
\textbf{Subspace-based Client Similarity.} FKPL uses the one-shot subspace comparison strategy of PACFL~\cite{DBLP:conf/aaai/VahidianMWK0S023}. Let $\mathcal{D}_{m,k}$ denote client $m$'s samples of class $k$. For each observed class, let $X_{m,k}\in\mathbb{R}^{d_x\times |\mathcal{D}_{m,k}|}$ contain vectorized local inputs as columns. The client retains leading left singular vectors from these matrices, with its basis-vector budget distributed across classes according to their local sample proportions. Let $Q_m$ collect the retained vectors and let $\mathcal{U}_m=\operatorname{span}(Q_m)$ denote the represented subspace. Using the spans makes the definition independent of the particular basis and allows vectors retained from different classes to be combined.

After receiving the client summaries, the server forms a dissimilarity matrix $A$ from the smallest principal angle between each pair of subspaces:
\begin{equation}
\label{eq:subspace_distance}
A_{ij}=\frac{180}{\pi}
\min_{\substack{\boldsymbol{u}\in\mathcal{U}_i,\,
                 \boldsymbol{v}\in\mathcal{U}_j\\
                 \|\boldsymbol{u}\|_2=\|\boldsymbol{v}\|_2=1}}
\arccos\!\left(|\boldsymbol{u}^{\top}\boldsymbol{v}|\right).
\end{equation}
The angles and clustering thresholds are expressed in degrees; smaller values indicate closer subspaces. This computation is performed once before expert training. Inputs are resized to $32\times32$, giving $d_x=3072$ for RGB images.

\textbf{Cluster Formation and Expert Training.} Average-linkage agglomerative clustering~\cite{day1984efficient} applied to $A$ gives a partition
\begin{equation}
\label{eq:partition}
\mathcal{I}=\{I_1,\ldots,I_V\},\qquad
\bigcup_{v=1}^{V} I_v=\{1,\ldots,M\},
\end{equation}
where the sets $I_v$ are disjoint and $V$ is the resulting number of clusters. This partition remains fixed during training. Within each cluster, FedAvg~\cite{conf/aistats/McMahanMRHA17} trains an expert using sample-size-weighted model aggregation. If $w_m^{t+1}$ is client $m$'s locally updated model in round $t$, the cluster model is updated as
\begin{equation}
\label{eq:cluster_aggregation}
w_v^{t+1}=\sum_{m\in I_v}
\frac{N_m}{\sum_{j\in I_v}N_j}\,w_m^{t+1}.
\end{equation}
The resulting expert parameters are $w_v^T=(\theta_v^T,\phi_v^T)$ after $T$ communication rounds.

\textbf{Global Integration by Feature Concatenation.} The global extractor retains the trained encoder from each cluster. Writing $\theta=(\theta_1^T,\ldots,\theta_V^T)$, its output is
\begin{equation}
\label{eq:global_features}
\boldsymbol{r}=\mathcal{E}(\boldsymbol{x};\theta)
=\mathcal{F}(\boldsymbol{x};\theta_1^T)\circ\cdots\circ
\mathcal{F}(\boldsymbol{x};\theta_V^T)\in\mathbb{R}^{dV},
\end{equation}
where $\circ$ denotes vector concatenation. Every expert's representation is retained as a distinct block, and its encoder parameters remain unchanged during integration and subsequent calibration. A linear classifier on this representation can reproduce any individual expert's linear prediction rule by using the corresponding block and assigning zero weights to the other blocks. It can also combine information from several experts. The resulting $K$-class head has $KdV+K$ trainable parameters, and the global extractor evaluates all $V$ encoders for each input.

\subsubsection{Prototype-Based Classifier Calibration}
PCC learns how to use the concatenated features through a server-trained linear classifier $\mathcal{C}(\cdot;\Phi):\mathbb{R}^{dV}\rightarrow\mathbb{R}^{K}$. Same-class domain prototypes provide reference points for constructing features that combine local sample information with representations from multiple clusters. The encoder parameters $\theta$ remain frozen throughout this stage.

\textbf{Prototype Aggregation.} Let $\mathcal{D}_{m,k}$ denote client $m$'s samples of class $k$, and let $\mathcal{K}_m=\{k:|\mathcal{D}_{m,k}|>0\}$. Using the broadcast global extractor, the client computes the local class prototype
\begin{equation}
\label{lcprtp}
\boldsymbol{p}_m^k=
\frac{1}{|\mathcal{D}_{m,k}|}
\sum_{(\boldsymbol{x},k)\in\mathcal{D}_{m,k}}
\mathcal{E}(\boldsymbol{x};\theta),
\qquad k\in\mathcal{K}_m.
\end{equation}
Define $I_{v,k}=\{m\in I_v:k\in\mathcal{K}_m\}$ as the clients in cluster $v$ that contain class $k$. For every nonempty $I_{v,k}$, the server forms an equal-client mean:
\begin{equation}
\label{dptpcmp}
\boldsymbol{s}_v^k=
\frac{1}{|I_{v,k}|}\sum_{m\in I_{v,k}}\boldsymbol{p}_m^k.
\end{equation}
Let $\mathcal{V}_k=\{v:|I_{v,k}|>0\}$ and
$\mathcal{S}^k=\{\boldsymbol{s}_v^k:v\in\mathcal{V}_k\}$.
The server broadcasts the collection $\mathcal{S}=\bigcup_{k=1}^{K}\mathcal{S}^k$.
Both $\boldsymbol{p}_m^k$ and $\boldsymbol{s}_v^k$ have dimension $dV$.

\textbf{Virtual Feature Synthesis.} For a local example $(\boldsymbol{x}_m^{k,i},k)$, the corresponding representation is
\begin{equation}
\label{exrfck}
\boldsymbol{r}_m^{k,i}=\mathcal{E}(\boldsymbol{x}_m^{k,i};\theta)
\in\mathbb{R}^{dV}.
\end{equation}
The representations computed for prototype estimation can be cached and reused for feature synthesis. Using an available same-class domain prototype, the client constructs
\begin{equation}
\label{svfvri}
\boldsymbol{g}_{m,v}^{k,i}
=(1-\alpha_{m,v}^{k,i})\boldsymbol{r}_m^{k,i}
+\alpha_{m,v}^{k,i}\boldsymbol{s}_v^k,
\qquad v\in\mathcal{V}_k,
\end{equation}
where $\alpha_{m,v}^{k,i}\sim U(a,b)$ and $0\leq a<b\leq1$.
The feature inherits the class label $k$. Client $m$ collects the resulting labelled features in $\mathcal{G}_m$ and uploads them once. This construction introduces same-class reference information from each represented cluster into calibration, while retaining the local sample contribution through the factor $1-\alpha_{m,v}^{k,i}$.

\textbf{Server-side Calibration.} The server pools the uploaded labelled features as $\mathcal{G}$ and minimizes their cross-entropy loss:
\begin{equation}
\label{eq:calibration_objective}
\mathcal{L}_{\mathrm{cal}}(\Phi)
=\frac{1}{|\mathcal{G}|}
\sum_{(\boldsymbol{g},y)\in\mathcal{G}}
\ell_{\mathrm{CE}}(\mathcal{C}(\boldsymbol{g};\Phi),y).
\end{equation}
For a minibatch $\mathcal{B}\subseteq\mathcal{G}$, the cross-entropy gradient is
\begin{equation}
\label{rcuvf}
\boldsymbol{g}_{\Phi}(\mathcal{B})=
\frac{1}{|\mathcal{B}|}
\sum_{(\boldsymbol{g},y)\in\mathcal{B}}
\nabla_{\Phi}\ell_{\mathrm{CE}}(\mathcal{C}(\boldsymbol{g};\Phi),y),
\end{equation}
The server uses this gradient with the SGD learning rate, momentum, and weight decay specified in \ref{app:hyperparameters}.
The server trains the head for $R$ epochs and broadcasts the calibrated classifier. The final prediction rule is
$\mathcal{C}(\mathcal{E}(\boldsymbol{x};\theta);\Phi)$.
After the one-time representation exchange, all classifier optimization takes place on the server. Algorithm~\ref{alg} gives the complete procedure.

\begin{algorithm}[t]
\small
\SetNoFillComment
\SetArgSty{textnormal}
\KwIn{Client datasets $\{\mathcal{D}_m\}_{m=1}^{M}$; expert-training rounds $T$; local epochs $E$; calibration epochs $R$; mixing bounds $a,b$; clustering and optimizer settings.}
\KwOut{Global extractor parameters $\theta$ and calibrated classifier parameters $\Phi$.}
\ForEach{client $m$ \textbf{in parallel}}{
  Extract and upload the retained local singular vectors $Q_m$\;
}
Compute $A$ using Eq.~\eqref{eq:subspace_distance} and obtain the fixed partition $\{I_v\}_{v=1}^{V}$\;
Initialize a cluster model $w_v^0$ for each $v$\;
\For{$t=0,\ldots,T-1$}{
  \ForEach{cluster $v$ \textbf{in parallel}}{
    Broadcast $w_v^t$ to clients in $I_v$\;
    \ForEach{client $m\in I_v$ \textbf{in parallel}}{
      Initialize $w_m\leftarrow w_v^t$\;
      Train $w_m$ on $\mathcal{D}_m$ for $E$ local epochs to obtain and upload $w_m^{t+1}$\;
    }
    Aggregate $w_v^{t+1}$ using Eq.~\eqref{eq:cluster_aggregation}\;
  }
}
Construct and freeze $\theta=(\theta_1^T,\ldots,\theta_V^T)$; broadcast the global extractor $\mathcal{E}$\;
\ForEach{client $m$ \textbf{in parallel}}{
  Extract local global-feature vectors and upload prototypes from Eq.~\eqref{lcprtp}\;
}
Aggregate domain prototypes using Eq.~\eqref{dptpcmp} and broadcast $\mathcal{S}$\;
\ForEach{client $m$ \textbf{in parallel}}{
  Synthesize labelled virtual features using Eq.~\eqref{svfvri} and upload $\mathcal{G}_m$\;
}
Pool the uploaded features as $\mathcal{G}$ and initialize a $dV$-input, $K$-output linear classifier $\Phi$\;
\For{$r=1,\ldots,R$}{
  Update $\Phi$ over minibatches of $\mathcal{G}$ with the specified optimizer and the gradient in Eq.~\eqref{rcuvf}\;
}
Broadcast the calibrated classifier $\Phi$\;
\caption{Federated Knowledge-Preserving Learning}
\label{alg}
\label{algO}
\end{algorithm}

\subsection{Threat Model and Privacy Evaluation Scope}
\textbf{Information Available to the Server.} We consider an honest-but-curious server that follows the protocol and can inspect client subspace summaries, uploaded expert parameters, local class prototypes, and labelled virtual features. It also knows the global extractor and the aggregated domain prototypes. Raw input samples remain on clients. The representation-inversion experiment in Section~\ref{sec:privacy_evaluation} measures reconstruction quality from the exchanged feature representations using the frozen global extractor.

\textbf{Information Available to a Participating Client.} An in-cluster adversary knows its own data and prototypes, the shared cluster model, the global extractor, and the broadcast domain prototypes. For a class present at more than one client in its cluster, it can subtract its own contribution from the equal-client aggregate to recover the average prototype of the other contributing clients. The client-side evaluation measures reconstruction from this derived statistic.

Subspace summaries, prototypes, and virtual features expose different levels of sample and class information. Prototype averaging and stochastic interpolation alter the reconstruction targets, and their effects are assessed through the specified inversion experiments. The reported reconstruction measurements characterize these observation and attack models.

\section{Experiments}
\label{experiments}

\subsection{Experimental Setup}
\textbf{Datasets and client allocation.} We evaluate FKPL on three multi-domain image-classification benchmarks. Office-Caltech-10 has ten classes in each of four domains: Caltech, Amazon, Webcam, and DSLR~\cite{DBLP:conf/cvpr/GongSSG12}. PACS contains seven classes across Photo, Art Painting, Cartoon, and Sketch~\cite{DBLP:conf/iccv/LiYSH17}. Domain-Net4 is a subset of DomainNet~\cite{9010750} formed by selecting ten classes and using Clipart, Painting, Real, and Sketch. Sampling up to 200 images per class and domain gives 7,914 images: 1,914 from Clipart and 2,000 from each remaining domain.

The experiments use 10 clients for Office-Caltech-10, 20 for PACS, and 10 for Domain-Net4. Each client draws its data from one domain. Domain-level training and test pools follow an 80/20 split. The domain allocations and local sampling ratios are specified in Table~\ref{tab:hyperparameter_training}. Client counts and sampling fractions vary across domains to represent different participation levels. The reported local sample counts range from 112 to 268 on Office-Caltech-10, from 98 to 214 on PACS, and from 400 to 1,600 on Domain-Net4, with means of 189, 156, and 618 samples per client, respectively. All methods use the same client allocation and held-out test pools for each benchmark.

\textbf{Models and comparisons.} The base model uses a ResNet-10 encoder~\cite{DBLP:conf/cvpr/HeZRS16} and a linear classifier, following the backbone configuration used in multi-domain FL studies~\cite{DBLP:conf/cvpr/HuangYSL023,DBLP:conf/cvpr/ChenHY24}. Each encoder outputs $d=512$ features. With $V=4$ experts, FKPL has a 2,048-dimensional concatenated representation. The classifier outputs seven logits for PACS and ten for the other datasets.

We compare FKPL with twelve published methods: FedAvg~\cite{conf/aistats/McMahanMRHA17}, FedProx~\cite{conf/mlsys/LiSZSTS20}, MOON~\cite{conf/cvpr/LiHS21}, FedProc~\cite{journals/fgcs/MuSCGFZZ23}, FedProto~\cite{conf/aaai/TanLLZ00Z22}, FPL~\cite{DBLP:conf/cvpr/HuangYSL023}, FedHEAL~\cite{DBLP:conf/cvpr/ChenHY24}, FedRC~\cite{DBLP:conf/pmlr/Guo24}, FedBR~\cite{DBLP:conf/pmlr/Guo23}, FedGMKD~\cite{NEURIPS2024_d6520fa7}, FDSE~\cite{11094889}, and FedConcat~\cite{FedConcat}. Two additional architecture controls examine encoder width and grouping. FedAvg-Wide uses a single encoder with output dimension $Vd$ and trains it using FedAvg. Ensemble-Independent retains the number and sizes of the FKPL groups and randomly reassigns clients to those groups.

\textbf{Training.} Federated training runs for $T=100$ rounds with full client participation, $E=10$ local epochs, and batch size 64. Local optimization uses SGD with learning rate 0.01, momentum 0.9, and weight decay $10^{-5}$. FKPL applies these settings within each cluster. Its one-shot average-linkage clustering returns four clusters for each benchmark, and the resulting partitions remain fixed across training runs. Inputs are resized to $32\times32$. PCC trains the linear classifier for $R=200$ epochs using the optimizer settings in \ref{app:hyperparameters}.

The reported classification results are averages over five independent training runs with different random seeds, using a fixed train/test split. Hyperparameters are selected using validation data. The final global predictor is evaluated on held-out samples from every participating domain. Tables~\ref{tab:hyperparameter_training} and~\ref{tab:hyperparameter_fkpl} summarize the shared training configuration and FKPL-specific settings.

\textbf{Evaluation metrics.} Let $A_i\in[0,100]$ be the reported accuracy, expressed as a percentage, on domain $i$, and let $D$ be the number of evaluated domains. The summary metrics are
\begin{equation}
\operatorname{AVG}=\frac{1}{D}\sum_{i=1}^{D} A_i,
\label{eq:avg}
\end{equation}
\begin{equation}
\operatorname{STD}=\sqrt{\frac{1}{D-1}\sum_{i=1}^{D}
\bigl(A_i-\operatorname{AVG}\bigr)^2}.
\label{eq:std}
\end{equation}
AVG gives equal weight to the evaluated domains, while STD describes their accuracy dispersion in percentage points. The tabulated summaries are calculated from the listed per-domain values. This cross-domain STD measures a different quantity from variability across repeated training runs. Accordingly, improvements in AVG and STD describe complementary aspects of performance.

\subsection{Comparison with Baseline Methods}
Tables~\ref{tab:acc}--\ref{tab:acc_domainnet} report the classification results. FKPL achieves average accuracies of 73.01\%, 67.00\%, and 64.51\% on Office-Caltech-10, PACS, and Domain-Net4, respectively. Relative to FedAvg, these values correspond to gains of 3.86, 8.46, and 1.35 percentage points. The gains have a clear domain structure: DSLR accuracy increases from 44.67\% to 76.67\%, and PACS Sketch accuracy increases from 50.55\% to 79.62\%. On Domain-Net4, FKPL raises the lowest domain accuracy from 50.98\% to 56.00\% while reducing the cross-domain STD from 10.21 to 7.50, the smallest value among the evaluated methods. These results show that preserving specialized encoders and calibrating their joint representation can strengthen domains that are poorly served by parameter averaging.

The architecture controls examine encoder width and client grouping. FKPL exceeds FedAvg-Wide by 0.52, 8.22, and 0.48 percentage points on the three benchmarks. Its gains over Ensemble-Independent are 5.74 points on PACS and 1.06 points on Domain-Net4. The larger gains on PACS connect the benefit of domain-oriented grouping and calibration to the stronger differences among its visual styles. The ablations below further separate the contributions of encoder construction and classifier training.

\begin{table}[tbp]
\centering
\small
\caption{Accuracy (\%) on Office-Caltech-10. AVG and STD are the mean and sample standard deviation across the listed domain accuracies. In the summary columns, boldface denotes the best value overall and underlining the best comparison value. The last two rows are architecture controls.}
\label{tab:acc}
\setlength{\tabcolsep}{5pt}
\renewcommand{\arraystretch}{1.12}
\begin{tabular}{lrrrrrr}
\toprule
Method & Caltech & Amazon & Webcam & DSLR & AVG $\uparrow$ & STD $\downarrow$ \\
\midrule
FedAvg & 69.02 & 79.79 & 83.10 & 44.67 & 69.15 & 17.39 \\
FedProx & 69.02 & 78.63 & 73.79 & 46.67 & 67.03 & 14.13 \\
MOON & 68.13 & 79.16 & 79.31 & 45.33 & 67.98 & 15.98 \\
FedProc & 66.16 & 77.79 & 71.38 & 40.67 & 64.00 & 16.26 \\
FedProto & 69.11 & 79.68 & 84.48 & 57.33 & 72.65 & 12.06 \\
FPL & 73.57 & 81.16 & 82.76 & 48.00 & 71.37 & 16.09 \\
FedHEAL & 63.75 & 63.37 & 68.62 & 52.00 & 61.94 & 7.04 \\
FedRC & 70.62 & 79.79 & 75.86 & 52.66 & 69.73 & 11.99 \\
FedBR & 76.52 & 84.84 & 76.21 & 84.00 & \textbf{\underline{80.39}} & \textbf{\underline{4.66}} \\
FedGMKD & 34.44 & 37.50 & 78.62 & 74.84 & 56.35 & 23.62 \\
FDSE & 65.98 & 79.05 & 34.14 & 44.00 & 55.79 & 20.43 \\
FedConcat & 27.14 & 16.53 & 50.00 & 70.00 & 40.92 & 23.89 \\
\midrule
\rowcolor{lightgray}
FKPL & 63.84 & 77.37 & 74.14 & 76.67 & 73.01 & 6.27 \\
\midrule
FedAvg-Wide & 72.32 & 77.05 & 87.24 & 53.33 & 72.49 & 14.21 \\
Ensemble-Independent & 58.48 & 78.42 & 82.76 & 80.00 & 74.92 & 11.10 \\
\bottomrule
\end{tabular}
\end{table}

\begin{table}[tbp]
\centering
\small
\caption{Accuracy (\%) on PACS. AVG and STD are the mean and sample standard deviation across the listed domain accuracies. In the summary columns, boldface denotes the best value overall and underlining the best comparison value. The last two rows are architecture controls.}
\label{tab:acc_pacs}
\setlength{\tabcolsep}{5pt}
\renewcommand{\arraystretch}{1.12}
\begin{tabular}{lrrrrrr}
\toprule
Method & Photo & Art painting & Cartoon & Sketch & AVG $\uparrow$ & STD $\downarrow$ \\
\midrule
FedAvg & 64.37 & 49.85 & 69.40 & 50.55 & 58.54 & 9.85 \\
FedProx & 64.01 & 49.36 & 70.09 & 50.75 & 58.55 & 10.14 \\
MOON & 65.63 & 47.25 & 70.43 & 53.50 & 59.20 & 10.69 \\
FedProc & 71.80 & 50.93 & 77.18 & 64.61 & 66.13 & 11.37 \\
FedProto & 64.61 & 45.15 & 68.08 & 48.97 & 56.70 & 11.33 \\
FPL & 66.47 & 51.33 & 72.91 & 56.69 & 61.85 & 9.68 \\
FedHEAL & 65.15 & 44.17 & 55.73 & 46.90 & 52.99 & 9.49 \\
FedRC & 68.32 & 50.59 & 73.63 & 57.27 & 62.45 & 10.44 \\
FedBR & 73.71 & 52.11 & 68.03 & 65.45 & 64.83 & \textbf{\underline{9.15}} \\
FedGMKD & 40.42 & 27.57 & 53.83 & 28.25 & 37.52 & 12.37 \\
FDSE & 67.60 & 51.47 & 80.30 & 84.87 & \textbf{\underline{71.06}} & 14.96 \\
FedConcat & 33.29 & 19.02 & 35.60 & 77.76 & 41.42 & 25.31 \\
\midrule
\rowcolor{lightgray}
FKPL & 69.46 & 45.83 & 73.08 & 79.62 & 67.00 & 14.72 \\
\midrule
FedAvg-Wide & 65.03 & 46.72 & 73.27 & 50.11 & 58.78 & 12.51 \\
Ensemble-Independent & 44.85 & 69.44 & 67.96 & 62.80 & 61.26 & 11.31 \\
\bottomrule
\end{tabular}
\end{table}

\begin{table}[tbp]
\centering
\small
\caption{Accuracy (\%) on Domain-Net4. AVG and STD are the mean and sample standard deviation across the listed domain accuracies. In the summary columns, boldface denotes the best value overall and underlining the best comparison value. The last two rows are architecture controls.}
\label{tab:acc_domainnet}
\setlength{\tabcolsep}{5pt}
\renewcommand{\arraystretch}{1.12}
\begin{tabular}{lrrrrrr}
\toprule
Method & Clipart & Painting & Real & Sketch & AVG $\uparrow$ & STD $\downarrow$ \\
\midrule
FedAvg & 66.52 & 59.99 & 75.13 & 50.98 & 63.16 & 10.21 \\
FedProx & 67.58 & 59.64 & 75.04 & 51.52 & 63.45 & 10.14 \\
MOON & 66.17 & 58.85 & 76.13 & 50.67 & 62.96 & 10.83 \\
FedProc & 67.40 & 60.82 & 74.77 & 50.96 & 63.49 & 10.11 \\
FedProto & 66.23 & 59.47 & 74.53 & 50.51 & 62.69 & 10.19 \\
FPL & 68.03 & 59.55 & 74.49 & 50.72 & 63.20 & 10.33 \\
FedHEAL & 63.12 & 54.08 & 73.51 & 45.23 & 58.99 & 12.13 \\
FedRC & 69.86 & 62.00 & 77.38 & 52.55 & 65.45 & 10.65 \\
FedBR & 67.05 & 62.77 & 78.80 & 53.75 & \textbf{\underline{65.59}} & 10.40 \\
FedGMKD & 55.76 & 57.25 & 72.75 & 41.25 & 56.75 & 12.88 \\
FDSE & 57.11 & 51.06 & 67.96 & 38.20 & 53.58 & 12.41 \\
FedConcat & 54.97 & 56.05 & 74.75 & 37.85 & 55.91 & 15.08 \\
\midrule
\rowcolor{lightgray}
FKPL & 67.02 & 61.50 & 73.50 & 56.00 & 64.51 & \textbf{7.50} \\
\midrule
FedAvg-Wide & 69.53 & 58.80 & 74.90 & 52.90 & 64.03 & 9.99 \\
Ensemble-Independent & 69.20 & 56.67 & 74.33 & 53.58 & 63.45 & \underline{9.91} \\
\bottomrule
\end{tabular}
\end{table}

\subsection{Component and Hyperparameter Analysis}
\subsubsection{Prototype Mixing Strength}
\begin{figure}[tbp]
\centering
\begin{minipage}[b]{0.49\linewidth}
    \centering
    \includegraphics[trim=8 10 60 45, clip, width=\linewidth]{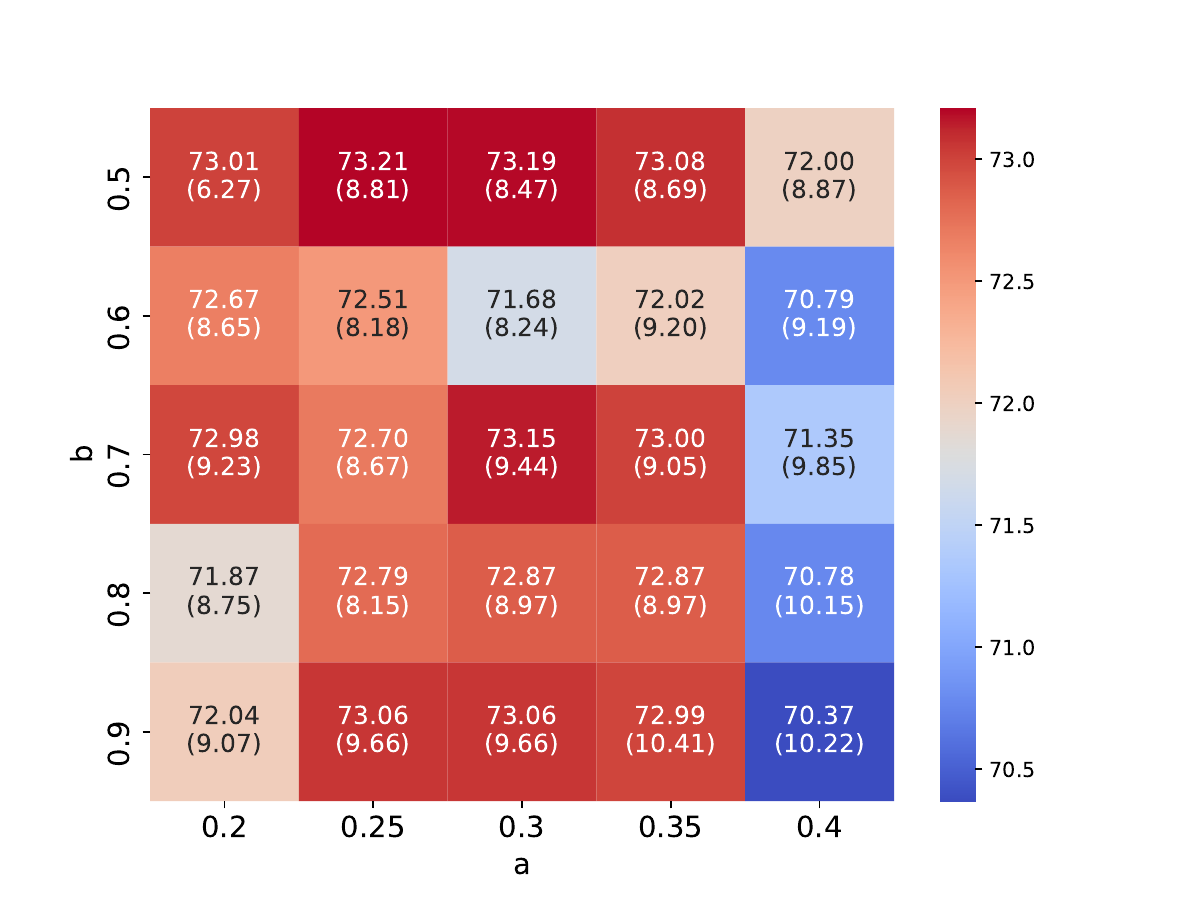}
    \captionsetup{font=small}
    \caption*{\small{(a) Office-Caltech-10}}
\end{minipage}
\hfill~
\begin{minipage}[b]{0.49\linewidth}
    \centering
\includegraphics[trim=8 10 60 45, clip, width=\linewidth]{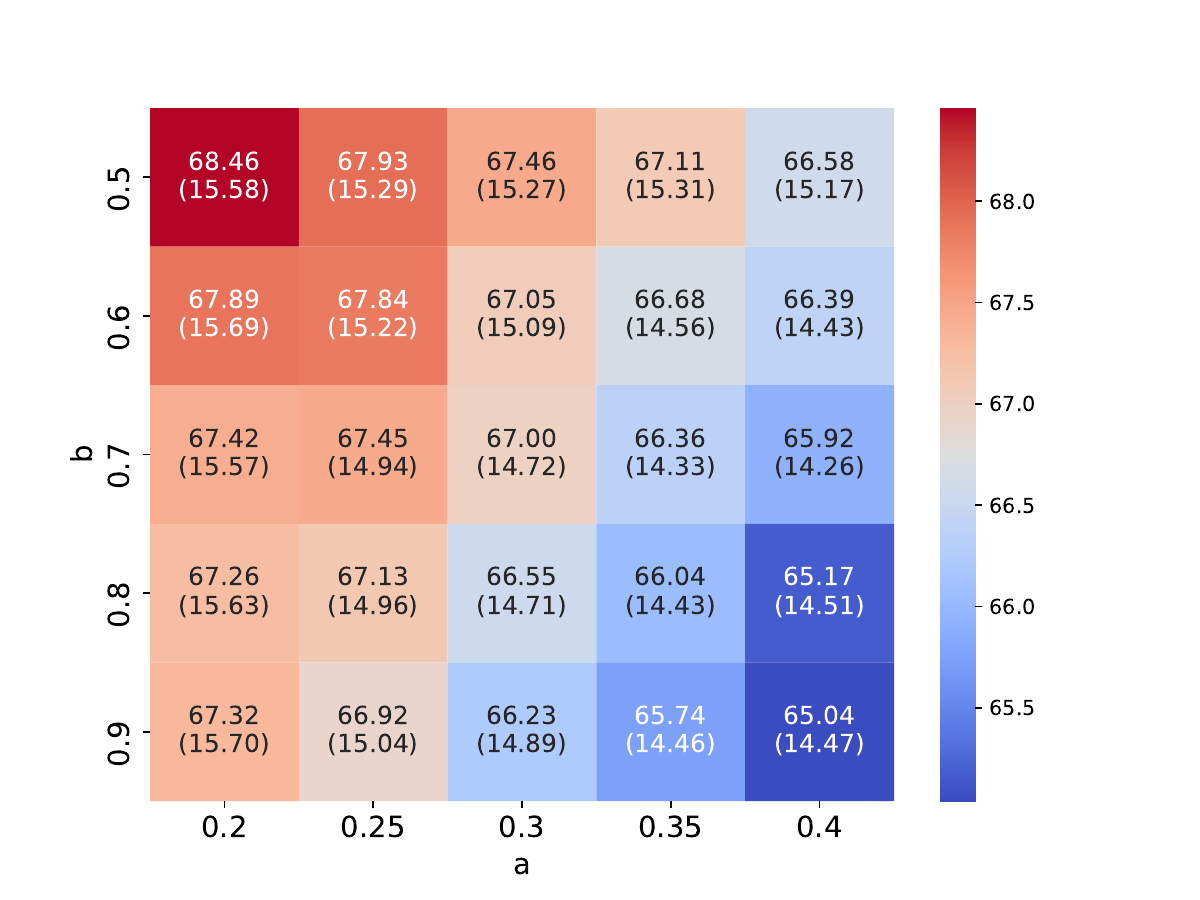}
    \captionsetup{font=small}
    \caption*{\small{(b) PACS}}
\end{minipage}
\caption{Sensitivity to the mixing bounds $a$ and $b$ in Eq.~\eqref{svfvri}. Each cell gives AVG with cross-domain STD in parentheses.}
\label{fig:hyperparam_study}

\end{figure}

Figure~\ref{fig:hyperparam_study} reports the effect of the mixing bounds $a$ and $b$ in Eq.~\eqref{svfvri}. On Office-Caltech-10, $(a,b)=(0.2,0.5)$ gives AVG 73.01\% and the smallest STD, 6.27, among the displayed settings; its AVG is close to the largest displayed value, 73.21\%. On PACS, increasing $a$ generally decreases both AVG and STD. A larger prototype contribution therefore changes the balance between mean accuracy and domain dispersion. We use $(0.3,0.7)$ for PACS and $(0.2,0.5)$ for Office-Caltech-10 and Domain-Net4.

\subsubsection{Contributions of DETFA and PCC}
The ablations isolate classifier calibration and encoder construction. With the concatenated extractor fixed, replacing federated classifier training with PCC improves AVG from 63.02\% to 73.01\% on Office-Caltech-10 and from 52.06\% to 67.00\% on PACS, yielding gains of 9.99 and 14.94 percentage points (Table~\ref{tab:ablation}). The corresponding DSLR and Sketch gains are 40.00 and 34.40 points. This concentration of improvements indicates that PCC is particularly useful when the classifier underuses representations from specific domains.

\begin{figure}[tbp]
\centering
\begin{minipage}[b]{0.49\linewidth}
    \centering
    \includegraphics[width=\linewidth]{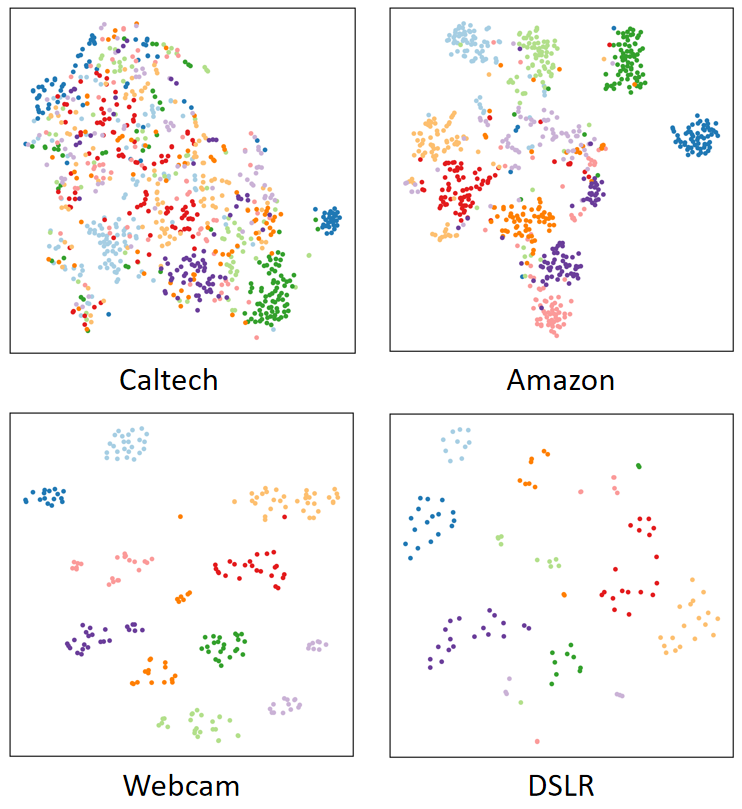}
    \captionsetup{font=small}
    \caption*{\small{(a) Office-Caltech-10}}
\end{minipage}
\hfill~
\begin{minipage}[b]{0.49\linewidth}
    \centering
    \includegraphics[width=\linewidth]{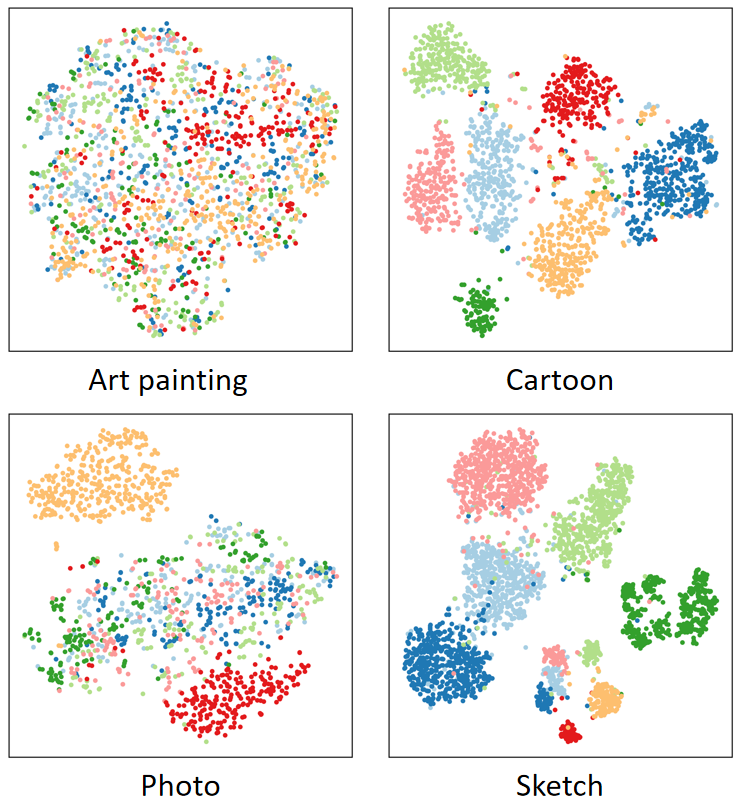}
    \captionsetup{font=small}
    \caption*{\small{(b) PACS}}
\end{minipage}

\caption{t-SNE visualizations of FKPL features on Office-Caltech-10 and PACS. Each panel shows features from the participating domains.}
\label{fig:tsne}
\end{figure}

With PCC retained, replacing the FedAvg encoder with DETFA increases AVG by 2.13 points on Office-Caltech-10 and 4.09 points on PACS (Table~\ref{tab:ablationDETFA}). The two experiments support complementary roles for specialized representation learning and prototype-guided calibration. Figure~\ref{fig:tsne} provides a qualitative view of the concatenated feature distributions.

\begin{table}[tbp]
\caption{Ablation of classifier training with the DETFA concatenated extractor fixed.}

\centering
\small 
\setlength{\tabcolsep}{2pt} 
\scalebox{1}{
\renewcommand\arraystretch{1.2}
\begin{tabular}{c||cccc|cc}
\hline 
\rowcolor{lightgray}
& \multicolumn{6}{c}{Office-Caltech-10}\\
\cline{2-7} 
\rowcolor{lightgray}
\multirow{-2}{*}{Methods}  
& Caltech & Amazon & Webcam & DSLR & AVG $\pmb{\uparrow}$ & STD $\pmb{\downarrow}$\\
\hline
Federated head
& 63.39 & 77.89 & 74.14 & 36.67 & 63.02 & 18.61 \\

\textbf{PCC} (proposed)
& 63.84 & 77.37 & 74.14 & 76.67 & \textbf{73.01} & \textbf{6.27} \\
\hline\thickhline 
\rowcolor{lightgray}
& \multicolumn{6}{c}{PACS}\\
\cline{2-7} 
\rowcolor{lightgray}
\multirow{-2}{*}{Methods}  
& Photo & Art painting & Cartoon & Sketch & AVG $\pmb{\uparrow}$ & STD $\pmb{\downarrow}$\\
\hline

Federated head
& 58.08 & 40.20 & 64.74 & 45.22 & 52.06 & \textbf{11.32} \\

\textbf{PCC} (proposed)
& 69.46 & 45.83 & 73.08 & 79.62 & \textbf{67.00} & 14.72 \\
\hline
\end{tabular}
}
\label{tab:ablation}

\end{table}

\begin{table}[tbp]
\caption{Ablation of encoder construction with PCC used for classifier calibration.}

\centering
\small
\setlength{\tabcolsep}{2pt}
\scalebox{1}{
\renewcommand\arraystretch{1.2}
\begin{tabular}{c||cccc|cc}
\hline
\rowcolor{lightgray}
& \multicolumn{6}{c}{Office-Caltech-10}\\
\cline{2-7}
\rowcolor{lightgray}
\multirow{-2}{*}{Methods}
& Caltech & Amazon & Webcam & DSLR & AVG $\pmb{\uparrow}$ & STD $\pmb{\downarrow}$\\
\hline
FedAvg encoder
& 70.98 & 80.00 & 75.86 & 56.67 & 70.88 & 10.16 \\

\textbf{DETFA}~(proposed)
& 63.84 & 77.37 & 74.14 & 76.67 & \textbf{73.01} & \textbf{6.27} \\
\hline\thickhline
\rowcolor{lightgray}
& \multicolumn{6}{c}{PACS}\\
\cline{2-7}
\rowcolor{lightgray}
\multirow{-2}{*}{Methods}
& Photo & Art painting & Cartoon & Sketch & AVG $\pmb{\uparrow}$ & STD $\pmb{\downarrow}$\\
\hline
FedAvg encoder
& 70.66 & 44.74 & 71.79 & 64.46 & 62.91 & \textbf{12.54} \\

\textbf{DETFA}~(proposed)
& 69.46 & 45.83 & 73.08 & 79.62 & \textbf{67.00} & 14.72 \\
\hline
\end{tabular}
}
\label{tab:ablationDETFA}

\end{table}

\subsubsection{Personalized Classifiers}
\begin{figure}[tbp]
\centering
\begin{minipage}[b]{0.48\linewidth}
    \includegraphics[trim=8 10 10 7, clip, width=\linewidth]{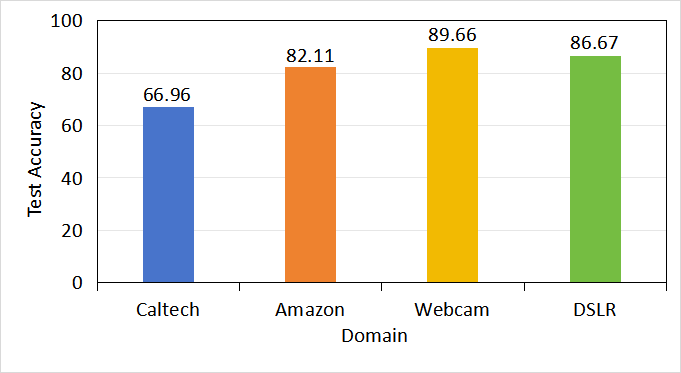}
    
    \captionsetup{font=small}
    \caption*{\small{(a) Office-Caltech-10}}
    
\end{minipage}
~
\begin{minipage}[b]{0.48\linewidth}
    \includegraphics[trim=8 10 10 7, clip, width=\linewidth]{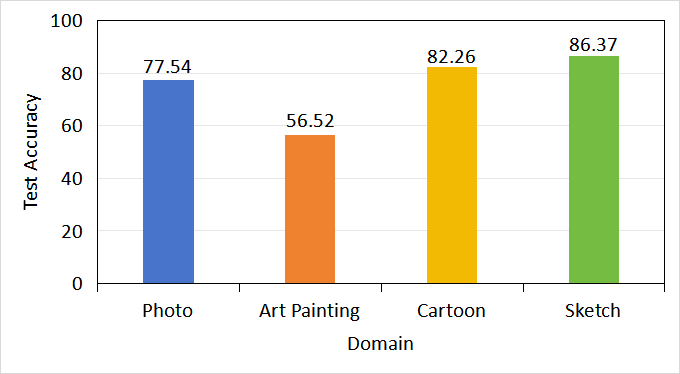}
    
    \captionsetup{font=small}
    \caption*{\small{(b) PACS}}
    
\end{minipage}

\caption{Accuracy of the cluster-specific classifiers trained on the frozen FKPL representation.}
\label{fig:personalization}

\end{figure}

The frozen concatenated extractor also supports cluster-specific prediction heads. Clients in the same cluster can jointly train a classifier on their local samples using FedAvg while retaining the shared extractor. This follows the personalized FL objective of adapting prediction to a client's data distribution~\cite{conf/mm/YiWLSY23,journals/tkde/LiSLLSYZ24}. Figure~\ref{fig:personalization} reports the resulting domain accuracies. Each cluster-specific head improves the corresponding domain accuracy relative to the unified FKPL head. The same collection of preserved features therefore supports both a common prediction rule and classifiers adapted to individual clusters.

\subsection{Representation Inversion Evaluation}
\label{sec:privacy_evaluation}
We examine reconstruction from four representations: raw features, local prototypes, aggregated prototypes, and virtual features. For each target, the attacker optimizes a $32\times32$ image by matching the target representation through the frozen global extractor. Adam runs for 800 steps with learning rate 0.05, total-variation regularization weight $10^{-5}$, and $L_2$ regularization weight $10^{-6}$. Table~\ref{tab:privacy_combined} reports PSNR, SSIM, and the final feature-matching loss. Lower PSNR and SSIM indicate lower similarity to the reference image under this reconstruction procedure.

Virtual features yield the lowest PSNR and SSIM among the four tested representations on both benchmarks. Relative to raw features, PSNR decreases from 8.43 to 6.92 dB on Office-Caltech-10 and from 9.13 to 7.58 dB on PACS. SSIM decreases from 0.2073 to 0.1685 and from 0.2347 to 0.2169, respectively. These comparisons quantify the effect of prototype-guided interpolation on the direct representation-matching attack. Prototype inversion measures reconstruction from class-level averages, so its target carries information aggregated across several examples.

We also evaluate an adversarial client that derives a representation of the other clients in its cluster. For a class $k$ with $|I_{v,k}|>1$, client $a\in I_{v,k}$ subtracts its own prototype from the equal-client aggregate:
\begin{equation}
  \widehat{\boldsymbol p}_{\mathrm{others}}^k
  =\frac{|I_{v,k}|\boldsymbol s_v^k-\boldsymbol p_a^k}
  {|I_{v,k}|-1}.
  \label{eq:others_prototype}
\end{equation}
This expression exactly recovers the mean prototype of the other contributing clients; for two contributing clients, it recovers the other client's local prototype. The attacker applies the same inversion procedure to this derived target. Table~\ref{tab:privacy_malicious_combined} reports the reconstruction results. Subtracting the client's own contribution lowers PSNR on Office-Caltech-10 and raises it on PACS, showing that the reconstruction effect depends on the target statistic and dataset. The two evaluations characterize distinct information available through virtual-feature exchange and shared domain prototypes.

\begin{table}[tbp]
\caption{Representation-inversion results on Office-Caltech-10 and PACS. PSNR is measured in dB; lower PSNR and SSIM indicate lower reconstruction similarity under the evaluated attack. Feature loss measures the representation-matching objective.}

\centering
\small
\setlength{\tabcolsep}{2pt}
\scalebox{1}{
\renewcommand\arraystretch{1.2}
\begin{tabular}{c|ccc}
\hline
\rowcolor{lightgray}
& \multicolumn{3}{c}{Office-Caltech-10}\\
\cline{2-4}
\rowcolor{lightgray}
\multirow{-2}{*}{Representation}
& PSNR$\downarrow$ & SSIM$\downarrow$ & Feature Loss\\
\hline
Raw feature
& 8.43 $\pm$ 2.21 & 0.2073 $\pm$ 0.0368 & 0.0543 $\pm$ 0.0392 \\
Local prototype
& 9.52 $\pm$ 2.09 & 0.2087 $\pm$ 0.0436 & 0.0444 $\pm$ 0.0237 \\
Aggregated prototype
& 9.19 $\pm$ 0.60 & 0.1973 $\pm$ 0.0187 & 0.0410 $\pm$ 0.0100 \\
Virtual feature
& \textbf{6.92 $\pm$ 1.39} & \textbf{0.1685 $\pm$ 0.0243} & 0.0481 $\pm$ 0.0237 \\
\hline\thickhline
\rowcolor{lightgray}
& \multicolumn{3}{c}{PACS}\\
\cline{2-4}
\rowcolor{lightgray}
\multirow{-2}{*}{Representation}
& PSNR$\downarrow$ & SSIM$\downarrow$ & Feature Loss\\
\hline
Raw feature
& 9.13 $\pm$ 3.31 & 0.2347 $\pm$ 0.0808 & 0.0308 $\pm$ 0.0327 \\
Local prototype
& 9.00 $\pm$ 3.18 & 0.2288 $\pm$ 0.0430 & 0.0330 $\pm$ 0.0294 \\
Aggregated prototype
& 10.07 $\pm$ 0.28 & 0.2489 $\pm$ 0.0107 & 0.0174 $\pm$ 0.0024 \\
Virtual feature
& \textbf{7.58 $\pm$ 1.46} & \textbf{0.2169 $\pm$ 0.0368} & 0.0219 $\pm$ 0.0097 \\
\hline
\end{tabular}
}
\label{tab:privacy_combined}

\end{table}

\begin{table}[tbp]
\caption{Representation inversion by an adversarial in-cluster client. The derived target is the mean prototype of the other contributing clients in Eq.~\eqref{eq:others_prototype}.}

\centering
\small
\setlength{\tabcolsep}{2pt}
\scalebox{1}{
\renewcommand\arraystretch{1.2}
\begin{tabular}{c|ccc}
\hline
\rowcolor{lightgray}
& \multicolumn{3}{c}{Office-Caltech-10}\\
\cline{2-4}
\rowcolor{lightgray}
\multirow{-2}{*}{Representation}
& PSNR$\downarrow$ & SSIM$\downarrow$ & Feature Loss\\
\hline
Shared cluster prototype
& 9.56 $\pm$ 2.09 & 0.2154 $\pm$ 0.0389 & 0.0397 $\pm$ 0.0207 \\
Leave-one-out others prototype
& \textbf{8.62 $\pm$ 1.84} & \textbf{0.2013 $\pm$ 0.0308} & 0.0532 $\pm$ 0.0280 \\
\hline\thickhline
\rowcolor{lightgray}
& \multicolumn{3}{c}{PACS}\\
\cline{2-4}
\rowcolor{lightgray}
\multirow{-2}{*}{Representation}
& PSNR$\downarrow$ & SSIM$\downarrow$ & Feature Loss\\
\hline
Shared cluster prototype
& \textbf{10.19 $\pm$ 3.15} & \textbf{0.2452 $\pm$ 0.0516} & 0.0234 $\pm$ 0.0253 \\
Leave-one-out others prototype
& 11.31 $\pm$ 2.58 & 0.2562 $\pm$ 0.0309 & 0.0136 $\pm$ 0.0135 \\
\hline
\end{tabular}
}
\label{tab:privacy_malicious_combined}

\end{table}

\subsection{Communication and Runtime}
\textbf{Communication accounting.} Table~\ref{tab:one_shot_comm_PCC} counts the model-parameter and representation scalars exchanged by each component of the protocol. Let $P_f$ denote the parameter count of one encoder, let $P_w=P_f+Kd+K$ be the parameter count of one expert including its local classifier, let $p_m$ be client $m$'s number of uploaded subspace vectors, and let $N_g$ be the total number of uploaded virtual features. Counts assume full participation and direct delivery to each receiving client, and exclude labels and message metadata. Each transmitted scalar uses four bytes under FP32 storage.

This accounting includes delivery of the concatenated encoder, local-prototype uploads, the complete domain-prototype broadcast, virtual-feature uploads, and the final classifier broadcast. When every cluster contains every class, the full prototype collection has $VKdV$ scalar entries per receiving client. The size of the virtual-feature upload depends on $N_g$ and the feature dimension $dV$. Following these exchanges, the server can run additional calibration epochs without another client upload. The communication benefit of PCC therefore comes from separating iterative classifier optimization from feature collection.

\textbf{Profiling configuration.} The measurements in Tables~\ref{tab:fkpl_comm_runtime_fused}--\ref{tab:inference_profile} were obtained on a workstation running Windows 11 Professional with an AMD Ryzen 9 9950X 16-core CPU, 128 GB RAM, and an NVIDIA GeForce RTX 4090 with 24 GB memory. The NVIDIA driver version was 560.94 and the CUDA version was 12.6. The reported training profiles use 10 Office-Caltech-10 clients and 20 PACS clients, matching the classification configurations.

\textbf{Training time.} Table~\ref{tab:fkpl_comm_runtime_fused} separates clustering, expert training, and server-side classifier calibration. Expert training takes 766.930 s on Office-Caltech-10 and 8,024.443 s on PACS. Server calibration takes 4.424 s and 21.437 s, respectively. With 200 calibration epochs, these correspond to approximately 0.022 s and 0.107 s per epoch. The measurements identify repeated expert optimization as the main recorded training cost and show that fitting the linear head contributes a comparatively small server workload.

\textbf{Memory and energy.} Table~\ref{tab:fkpl_profile_summary} reports the measured stage-specific GPU memory and energy values. Expert-stage energy is given per training round, whereas classifier-stage energy covers its one-time execution. With 100 expert-training rounds, expert optimization accounts for most of the recorded training energy. These measurements provide a resource profile for the concatenated architecture on the specified workstation.

\textbf{Inference.} Table~\ref{tab:inference_profile} compares a single expert with the concatenated extractor on the RTX 4090. At batch size one, the concatenated extractor takes 3.093 ms on PACS and 8.223 ms on Office-Caltech-10, compared with 0.795 ms and 0.801 ms for a single expert. At batch size 32, its per-sample latency is 0.099 ms and 0.472 ms, respectively. Batching amortizes execution overhead across samples. These measurements quantify the deployment cost of evaluating all experts on the tested GPU and motivate expert selection or compressed representations for deployments with tighter inference budgets.

\begin{table}[tbp]
\centering
\small
\caption{Analytical communication counts in transmitted scalar values. Expert training includes $T$ rounds; all remaining entries are one-time exchanges. Labels and message metadata are excluded. Each client receives the full concatenated encoder and complete available domain-prototype collection.}
\label{tab:one_shot_comm_PCC}
\renewcommand{\arraystretch}{1.2}
\begin{tabularx}{\linewidth}{@{}Xl@{}}
\toprule
Protocol component & Scalar count over all clients \\
\midrule
Subspace-summary uploads & $d_x\sum_{m=1}^{M}p_m$ \\
Expert-model uploads and downloads & $2TMP_w$ \\
Concatenated-encoder downloads & $MVP_f$ \\
Local-prototype uploads & $dV\sum_{m=1}^{M}|\mathcal K_m|$ \\
Domain-prototype downloads & $MdV\sum_{k=1}^{K}|\mathcal V_k|$ \\
Virtual-feature uploads & $N_gdV$ \\
Calibrated-classifier downloads & $M(KdV+K)$ \\
\bottomrule
\end{tabularx}
\end{table}

\begin{table}[tbp]
\centering
\small
\caption{Recorded runtime at the classification client scales. Stage I uses 100 communication rounds and Stage II uses 200 classifier-calibration epochs. Server calibration time covers optimization of the linear head.}
\label{tab:fkpl_comm_runtime_fused}
\setlength{\tabcolsep}{6pt}
\renewcommand{\arraystretch}{1.15}
\begin{tabularx}{\linewidth}{@{}Xrr@{}}
\toprule
Recorded operation & Office-Caltech-10 & PACS \\
 & 10 clients & 20 clients \\
\midrule
One-shot preprocessing time (s) & 0.436 & 7.309 \\
Server hierarchical-clustering time (s) & 0.006 & 0.015 \\
Stage I: total runtime (s) & 766.930 & 8,024.443 \\
Stage I: mean runtime per round (s) & 7.669 & 80.244 \\
Stage II: server calibration time (s) & 4.424 & 21.437 \\
Stage II: mean calibration time per epoch (s) & 0.022 & 0.107 \\
Evaluation time (s) & 0.732 & 1.318 \\
\bottomrule
\end{tabularx}
\end{table}

\begin{table}[tbp]
\centering
\small
\caption{Recorded memory and energy at the classification client scales. Encoder energy is measured per expert-training round; classifier-stage energy is measured for its one-time execution.}
\label{tab:fkpl_profile_summary}
\setlength{\tabcolsep}{6pt}
\renewcommand{\arraystretch}{1.15}
\begin{tabularx}{\linewidth}{@{}Xrr@{}}
\toprule
Metric & Office-Caltech-10 & PACS \\
 & 10 clients & 20 clients \\
\midrule
Peak GPU memory: encoder stage (MB) & 172.825 & 276.365 \\
Peak GPU memory: classifier stage (MB) & 133.022 & 103.481 \\
Peak GPU memory: evaluation (MB) & 329.966 & 623.595 \\
Mean encoder-round energy (J) & 180.680 & 3,082.015 \\
One-time classifier-stage energy (J) & 149.857 & 520.764 \\
\bottomrule
\end{tabularx}
\end{table}

\begin{table}[tbp]
\caption{Inference profiling on the RTX 4090. Batch latency and per-sample latency are shown separately; throughput is measured in samples per second. Peak memory and energy are reported for each profiling configuration.}

\centering
\small
\setlength{\tabcolsep}{3pt}
\makebox[\textwidth][c]{
\resizebox{0.95\textwidth}{!}{
\renewcommand\arraystretch{1.2}
\begin{tabular}{llcccccc}
\hline
\rowcolor{lightgray}
Dataset & Model & Batch
& Batch latency (ms)$\downarrow$
& Per-sample (ms)$\downarrow$
& Samples/s$\uparrow$
& Peak Mem. (MB)$\downarrow$
& Energy (J)$\downarrow$ \\
\hline
\multirow{4}{*}{PACS}
& Single expert & 1  & 0.795 & 0.795 & 1,257.70  & 126.27 & 0.0175 \\
& Single expert & 32 & 0.786 & 0.025 & 40,730.64 & 126.92 & 0.0260 \\
& concatenated extractor & 1  & 3.093 & 3.093 & 323.28   & 155.14 & 0.2864 \\
& concatenated extractor & 32 & 3.181 & 0.099 & 10,058.40 & 156.47 & 0.4436 \\
\hline
\multirow{4}{*}{Office-Caltech-10}
& Single expert & 1  & 0.801 & 0.801 & 1,248.72  & 254.23 & 0.0203 \\
& Single expert & 32 & 0.787 & 0.025 & 40,641.06 & 255.13 & 0.0406 \\
& concatenated extractor & 1  & 8.223 & 8.223 & 121.60   & 342.61 & 0.8951 \\
& concatenated extractor & 32 & 15.098 & 0.472 & 2,119.51  & 344.81 & 1.8482 \\
\hline
\end{tabular}
}}
\label{tab:inference_profile}

\end{table}

\subsection{Computational Complexity Analysis}
Let $N=\sum_{m=1}^{M}N_m$, let $d_x$ be the vectorized input dimension, and let $p$ bound the number of retained subspace vectors per client. Denote the cost of extracting client $m$'s singular vectors by $C_{\mathrm{SVD},m}$. For a dense SVD of a $d_x\times n$ class-data matrix, the arithmetic cost is $\mathcal{O}(d_xn\min\{d_x,n\})$; truncated solvers have solver-dependent costs. Principal-angle comparisons can be computed from orthonormal bases of the uploaded spans. Constructing these bases and all pairwise comparisons takes
\begin{equation}
\mathcal{O}\!\left(Md_xp^2+M^2(d_xp^2+p^3)\right)
\end{equation}
operations. If $C_{\mathrm{HC}}(M)$ denotes the cost of the chosen hierarchical-clustering implementation, the one-time clustering cost is
\begin{equation}
\mathcal{O}\!\left(
\sum_{m=1}^{M}C_{\mathrm{SVD},m}
+Md_xp^2+M^2(d_xp^2+p^3)+C_{\mathrm{HC}}(M)
\right).
\end{equation}
The dense client-dissimilarity matrix requires $\mathcal{O}(M^2)$ storage. The clustering term explicitly includes the agglomeration step in addition to pairwise distance construction.

Let $C_{\mathrm{train}}$ denote the per-sample forward and backward computation for a base expert. With full client participation, $T$ communication rounds, and $E$ local epochs, the total expert-training work is $\mathcal{O}(TEN C_{\mathrm{train}})$. This expression measures aggregate computation across clients; elapsed runtime also depends on client parallelism.

PCC follows expert training. Let $C_{\mathrm{enc}}$ be the per-sample inference cost of one encoder, so evaluating the concatenated extractor costs $\mathcal{O}(VC_{\mathrm{enc}})$. Let $N_g=|\mathcal{G}|$ be the number of uploaded virtual features, counted with their multiplicities. With cached local representations, feature extraction, prototype aggregation, feature synthesis, and $R$ epochs of linear-classifier training together cost
\begin{equation}
\mathcal{O}\!\left(
NVC_{\mathrm{enc}}+MKdV+N_gdV+RN_gKdV
\right).
\end{equation}
When each example is mixed with one prototype from every cluster, $N_g=NV$; missing class prototypes reduce this count according to the sets $\mathcal{V}_k$. The global head contains $KdV+K$ parameters. Inference evaluates $V$ encoders and one linear head, giving per-sample cost $\mathcal{O}(VC_{\mathrm{enc}}+KdV)$. These expressions distinguish the one-time calibration workload from the recurring cost of deploying the concatenated extractor.

\section{Conclusion}
\label{conclusion}
FKPL integrates domain-specialized encoders into a unified predictor through feature concatenation and prototype-based classifier calibration. Subspace-based grouping organizes expert training around data distributions, while virtual-feature synthesis supplies the server with calibration examples guided by domain prototypes. Experiments on Office-Caltech-10, PACS, and Domain-Net4 show improvements over FedAvg and identify classifier calibration as a major source of the gains. On Domain-Net4, FKPL also attains the lowest cross-domain accuracy dispersion among the evaluated methods. This combination of modular representation learning and server-side calibration provides a design for federated classification across participating domains. Future work will explore expert selection and representation compression to reduce deployment costs.

\clearpage
\appendix
\section{Detailed Hyperparameter Configurations}
\label{app:hyperparameters}

Tables~\ref{tab:hyperparameter_training} and~\ref{tab:hyperparameter_fkpl} summarize the experimental configurations.

\setcounter{table}{0}
\renewcommand{\thetable}{A\arabic{table}}
\renewcommand{\theHtable}{A.\arabic{table}}

\begin{table}[htbp]\small
\centering
    \caption{Federated training, model architecture, and client allocation.}
    \label{tab:hyperparameter_training}
    \resizebox{\linewidth}{!}{
    \renewcommand\arraystretch{1.3}
    \begin{tabular}{l||l|ccc}
    \hline
        \rowcolor{lightgray}
        Category & Hyperparameter & Office-Caltech-10 & PACS & Domain-Net4 \\
    \hline\hline

    \multirow{9}{*}{\makecell{Federated\\training}}
        & Communication rounds $T$ & 100 & 100 & 100 \\
        \cline{2-5}
        & Local epochs $E$ & 10 & 10 & 10 \\
        \cline{2-5}
        & Mini-batch size $B$ & 64 & 64 & 64 \\
        \cline{2-5}
        & Optimizer & SGD & SGD & SGD \\
        \cline{2-5}
        & Learning rate $\eta$ & 0.01 & 0.01 & 0.01 \\
        \cline{2-5}
        & Momentum & 0.9 & 0.9 & 0.9 \\
        \cline{2-5}
        & Weight decay & $1\times10^{-5}$ & $1\times10^{-5}$ & $1\times10^{-5}$ \\
        \cline{2-5}
        & Client participation ratio & 1.0 & 1.0 & 1.0 \\
        \cline{2-5}
        & Intra-cluster aggregation & \makecell{Sample-size weighted\\FedAvg} & \makecell{Sample-size weighted\\FedAvg} & \makecell{Sample-size weighted\\FedAvg} \\
    \hline

    \multirow{7}{*}{\makecell{Model\\architecture}}
        & Backbone & ResNet-10 & ResNet-10 & ResNet-10 \\
        \cline{2-5}
        & Basic channel width & 64 & 64 & 64 \\
        \cline{2-5}
        & Residual block configuration & $[1,1,1,1]$ & $[1,1,1,1]$ & $[1,1,1,1]$ \\
        \cline{2-5}
        & Feature dimension $d$ & 512 & 512 & 512 \\
        \cline{2-5}
        & Classifier & Single linear layer & Single linear layer & Single linear layer \\
        \cline{2-5}
        & Number of classes $K$ & 10 & 7 & 10 \\
        \cline{2-5}
        & Input image size & $32\times32$ & $32\times32$ & $32\times32$ \\
    \hline

    \multirow{4}{*}{\makecell{Client\\configuration}}
        & Total number of clients $M$ & 10 & 20 & 10 \\
        \cline{2-5}
        & Clients per domain
        & \makecell{Caltech: 4; Amazon: 3;\\ Webcam: 2; DSLR: 1}
        & \makecell{Photo: 5; Art Painting: 4;\\ Cartoon: 5; Sketch: 6}
        & \makecell{Sketch: 4; Clipart: 3;\\ Real: 2; Painting: 1} \\
        \cline{2-5}
        & Domain sampling ratio
        & \makecell{Caltech: 25\%; Amazon: 25\%;\\ Webcam: 50\%; DSLR: 100\%}
        & \makecell{Photo: 20\%; Art Painting: 20\%;\\ Cartoon: 20\%; Sketch: 10\%}
        & \makecell{Sketch: 25\%; Clipart: 30\%;\\ Real: 50\%; Painting: 100\%} \\
        \cline{2-5}
        & \makecell[l]{Concatenated feature\\dimension $dV$} & 2048 & 2048 & 2048 \\
    \hline

    \end{tabular}
    }
    
\end{table}

\begin{table}[htbp]\small
\centering
    \caption{Client clustering, prototype-based classifier calibration, and personalization.}
    \label{tab:hyperparameter_fkpl}
    \resizebox{\linewidth}{!}{
    \renewcommand\arraystretch{1.3}
    \begin{tabular}{l||l|ccc}
    \hline
        \rowcolor{lightgray}
        Category & Hyperparameter & Office-Caltech-10 & PACS & Domain-Net4 \\
    \hline\hline

    \multirow{7}{*}{Client clustering}
        & Subspace extraction & Truncated SVD & Truncated SVD & Truncated SVD \\
        \cline{2-5}
        & Client-level basis-vector budget & 10 & 10 & 10 \\
        \cline{2-5}
        & Similarity metric & Minimum principal angle $\theta_1$ & Minimum principal angle $\theta_1$ & Minimum principal angle $\theta_1$ \\
        \cline{2-5}
        & Clustering algorithm &  \makecell{Agglomerative\\hierarchical clustering}&  \makecell{Agglomerative\\hierarchical clustering} & \makecell{Agglomerative\\hierarchical clustering} \\
        \cline{2-5}
        & Linkage criterion & Average linkage & Average linkage & Average linkage \\
        \cline{2-5}
        & Clustering threshold (degrees) & 5.3 & 4.2 & 11.5 \\
        \cline{2-5}
        & Number of clusters $V$ & 4 & 4 & 4 \\
    \hline

    \multirow{6}{*}{\makecell{Prototype-based\\ classifier\\calibration}}
        & Mixing distribution $U(a,b)$ & $U(0.2,0.5)$ & $U(0.3,0.7)$ & $U(0.2,0.5)$ \\
        \cline{2-5}
        & Calibration epochs $R$ & 200 & 200 & 200 \\
        \cline{2-5}
        & Calibration batch size & 64 & 64 & 64 \\
        \cline{2-5}
        & Optimizer & SGD & SGD & SGD \\
        \cline{2-5}
        & Learning rate & 0.01 & 0.01 & 0.01 \\
        \cline{2-5}
        & Momentum / weight decay & $0.9$ / $1\times10^{-5}$ & $0.9$ / $1\times10^{-5}$ & $0.9$ / $1\times10^{-5}$ \\
    \hline

    \multirow{3}{*}{Personalization}
        & Classifier communication rounds & 100 & 100 & 100 \\
        \cline{2-5}
        & Local classifier epochs per round & 10 & 10 & 10 \\
        \cline{2-5}
        & Optimizer & SGD ($0.01$, $0.9$, $1\times10^{-5}$) & SGD ($0.01$, $0.9$, $1\times10^{-5}$) & SGD ($0.01$, $0.9$, $1\times10^{-5}$) \\
    \hline

    \end{tabular}
    }
\par\smallskip
\begin{minipage}{\linewidth}
\footnotesize

\noindent \textit{Note}: The \textit{client-level basis-vector budget} is set to 10 and distributed across classes according to local sample proportions. The resulting client partitions are fixed after the one-shot clustering procedure.
\end{minipage}
\end{table}

\clearpage
\bibliographystyle{elsarticle-num}
\bibliography{references}

\end{document}